%% file: Alterman2018-processed.tex
\documentclass[twocolumn]{aastex61}
\usepackage{xcolor}
\usepackage{bm}

\graphicspath{{./figs/}}
\definecolor{mygreen}{rgb}{0.02, 0.51, 0.13}

\input{preamble.tex}
\input{figs.tex}

\received{2018 April 5}
\revised{2018 July 6}
\accepted{2018 July 6}
\submitjournal{ApJ}
\shorttitle{Collisionless Solar Wind Differential Flow}
\shortauthors{Alterman et al.}
\begin{document}

\title{A comparison of Alpha Particle and Proton Beam Differential flow in Collisionally Young Solar Wind}

\input{affiliations.tex}

\begin{abstract}
In fast wind or when the local Coulomb collision frequency is low, observations show that solar wind minor ions and ion sub-populations flow with different bulk velocities.
Measurements indicate that the drift speed of both alpha particles and proton beams with respect to the bulk or core protons rarely exceeds the local Alfv\'en speed, suggesting that a magnetic instability or other wave-particle process limits their maximum drift.
We compare simultaneous alpha particle, proton beam, and proton core observations from instruments on the \emph{Wind} spacecraft spanning over 20 years.
In nearly collisionless solar wind, we find that the normalized alpha particle drift speed is slower than the normalized proton beam speed; no correlation between fluctuations in both species' drifts about their means; and a strong anti-correlation between collisional age and alpha-proton differential flow, but no such correlation with proton beam-core differential flow.
Controlling for the collisional dependence, both species' normalized drifts exhibit similar statistical distributions.
In the asymptotic, zero Coulomb collision limit, the youngest measured differential flows most strongly correlate with an approximation of the Alfv\'en speed that includes proton pressure anisotropy.
In this limit and with this most precise representation, alpha particles drift at 67\% and proton beam drift is approximately 105\% of the local Alfv\'en speed.
We posit that one of two physical explanations is possible.
Either (1) \replaced{a}{an} Alfv\'enic process preferentially accelerates or sustains proton beams and not alphas or (2) alpha particles are more susceptible to either an instability or Coulomb drag than proton beams.

\end{abstract}




%%%%%%%%%%
%%%%%%%%%%
%%%%%%%%%%
\section{Introduction} \label{sec:intro}
Simple models of solar wind acceleration (e.g.\ \citet{Parker1958a}) are unable to explain the solar wind's acceleration to high speeds.
Wave-particle interactions are likely necessary to explain these observations.
Differential flow is the velocity difference between two ion species. It is a useful indicator of such interactions and related acceleration.

Ionized hydrogen (protons) is the most common ion in the solar wind, usually constituting over 95\% by number density.
Within a few thermal widths of their mean speed, solar wind protons are well described by a single bi-Maxwellian velocity distribution function (VDF).
However, an asymmetric velocity space shoulder has also been observed in the proton distribution.
It can be described by a second, differentially flowing Maxwellian.
We refer to the primary proton component as the proton core ($p_1$) and the secondary component as the proton beam ($p_2$).
Proton beams are most easily measured in fast solar wind and when the local Coulomb collision frequency is small in comparison to the local expansion time.
Fully ionized helium (alpha particles, $\alpha$) are the second most common species and constitute $\sim 4\%$ of the solar wind by number density.

Differential flow is the velocity difference between two ion species or populations.
It has been measured in the solar wind plasma at many solar distances starting in the corona and, when the local collision rate is smaller than the expansion time, extending out to and beyond 1 AU.\citep{Landi2009,Marsch1982b,Marsch1982c,Steinberg1996,Neugebauer1976,Kasper2008,Feldman1974,Asbridge1976,Goldstein1995}
\Citet{Kasper2006} showed that $\alpha$ differential flow is aligned with the magnetic field $\BB$ to within several degrees as long as it is larger than $\sim 1\%$ of the measured solar wind speed, consistent with any apparent non-parallel flow being measurement error.
It should not be surprising that differential flow is field aligned because any finite differential flow perpendicular to $\BB$ would immediately experience a Lorentz force until the plasma was again gyrotropic on a timescale comparable to the ion gyroperiod.
We denote the differential flow as $\Delta v_{b,c} = \left(\Bv_b - \Bv_c\right) \cdot \bhat$, where ion species $b$ differentially streams with respect to core population $c$ and $\bhat$ is the magnetic field unit vector.
Positive differential flow is parallel to local $\BB$ and negative differential flow is antiparallel to it.
Simultaneous measurements of $\alpha$-particles and protons indicate that $\dv{{\alpha,p1}}$ is typically $\lesssim 70\%$ of the local Alfv\'en speed, $C_A$.\citep{Kasper2017,Kasper2008,Neugebauer1976,Asbridge1976,Feldman1974}
While measurements of heavier ions (e.g.\ iron, oxygen, carbon) show similar behavior \citep{Berger2011}, proton beam-core differential flow ($\dv{{p_2,p_1}}$) has been reported at approximately the local Alfv\'en speed or larger \citep{Marsch1982c}.
Given that the local Alfv\'en speed in the solar wind is generally a decreasing function of distance from the sun, this apparent Alfv\'en speed limit implies that there is effectively a local wave-mitigated limit on $\dv{{p2,p1}}$, for which several instability processes have been hypothesized. \citep{Daughton1998,*Daughton1999,*Goldstein2000}
Simulations by \citet{Maneva2015} showed that a nonlinear streaming instability limits alpha particle drift to a maximum of $0.5 \, C_A$.

Raw data from the Wind/SWE Faraday cups are now archived at the NASA Space Physics Data Facility (SPDF) and available online at CDAweb.
We have developed a new fitting algorithm that returns simultaneous parameters for three solar wind ion populations ($\alpha$, $p_1$, and $p_2$) and have processed over 20 years for Faraday cup solar wind measurements.
For this project, we have restricted the analysis to measurements with clear differential flow signatures for both the alpha particle and proton beam components.
We find that $\dv{{\alpha,p_1}}/C_A$ and $\dv{{p_2,p_1}}/C_A$ are indeed clustered around characteristic values that are consistent with previous results, but with considerable spreads in the respective distributions.
We investigate possible contributions to the spreads; the apparent impact of Coulomb collisions in the weakly-collisional regime; and the limitations of calculating the Alfv\'en speed under the commonly assumed frameworks of ideal and anisotropic MHD.
We report that in collisionless solar wind:
\begin{enumerate}
\item $\alpha$ particle and $p_2$ differential flow speeds exhibit distinctly different trends with the locally-measured Coulomb collision rate;
\item Coulomb collisions account for the dominant contribution to the spread in $\dv{}/C_A$;
\item and an accounting for the proton pressure anisotropy in the local Alfv\'en speed, as under anisotropic MHD, significantly reduces the spread in $\dv{}/C_A$.
\end{enumerate}

\noindent
For the most nearly collisionless solar wind measured at 1 AU and using the more precise, anisotropic approximation of the Alfv\'en speed we report that:
\begin{enumerate}
\item $\dv{{p_2,p_1}}$ is $106\%\pm 15\%$ of the local Alfv\'en speed;
\item $\dv{{\alpha,p_1}}$ is $62\%\pm 13\%$ of the local Alfv\'en speed;
\item and $\dv{{p_2,p_1}}\approx 1.7 \times \dv{{\alpha,p_1}}$.
\end{enumerate}

\noindent
Finally, we extrapolate to the perfectly collisionless limit, and estimate that:
\begin{enumerate}
\item $\dv{{p_2,p_1}}$ is $\sim 105\% \pm 15\%$ of the Alfv\'en speed and
\item $\dv{{\alpha,p_1}}$ is $67\% \pm 9\%$ of the Alfv\'en speed.
\end{enumerate}

\section{Data Sources \& Selection} \label{sec:data}
The \textit{Wind} spacecraft launched in fall 1994.
Its twin Faraday cup instruments have collected over 6.1 million proton and alpha particle direction-dependent energy spectra, the majority of which are in the solar wind.\citep{Ogilvie1995a}
Available on CDAweb, these raw spectra consist of measured charge flux as a function of angel and energy-per-charge for each cup.
With these spectra, we reconstruct 3D velocity distribution functions (VDFs) for each ion species and extract the bulk plasma properties: number density, velocity, and thermal speed.
Over more than 20 years, refinements in the data processing algorithms have yielded new information from these distributions including precise $\alpha$ particle abundances \citep{Aellig2001,Kasper2007,Kasper2012}, perpendicular to parallel proton temperature ratios \citep{Kasper2002b,Kasper2008}, and relative alpha to proton temperature ratios \citep{Kasper2008,Maruca2013}.

\Citet{Ogilvie1995a} provide a thorough description of the Solar Wind Experiment (SWE).
In summary, the SWE Faraday cups measure a single energy window approximately every $3\mathrm{s}$ and a full spectrum combines multiple energy windows measured over $\sim 92\mathrm{s}$.
Our fitting algorithm utilizes magnetic field measurements from the \textit{Wind} Magnetic Field Investigation (MFI) \citep{Koval2013a,Lepping1995} to determine each VDF's orientation relative to the local magnetic field and it assumes that the extracted parameters are approximately constant over the measurement time.
In spectra for which this is not the case, automatically processed bulk properties can be unreliable.

This new fitting algorithm returns 15 simultaneous parameters for three solar wind ion-populations: alpha particles ($\alpha$), proton cores ($p_1$) and proton beams ($p_2$).
\Citet{Kasper2006} describes the six parameter $\alpha$ fitting routines. 
The protons are jointly fit by a nine-parameter set: six to $p_1$ (number density, vector velocity, and parallel \& perpendicular temperature) and three to $p_2$ (number density, differential flow, and isotropic thermal speed).

Previous work with this data includes studies by \citet{Chen2016,*Gary2016}.
Figure\nobreakspace \ref {fig:thetaBn} shows example energy-per-charge measurements made in four representative look directions.
These directions are identified by the angle between the magnetic field and the direction normal to the Faraday cup's aperture.
Figure\nobreakspace \ref {fig:exVDFs} provides the corresponding proton (top) and $\alpha$ (bottom) VDFs.
The proton beam is the extension of the proton VDF to large $v_\parallel > 0$.

\plotThetaBn
\plotVDFex

Our alpha particle and proton core quality requirements nominally follow \citet{Kasper2002b,Kasper2007,Kasper2008}.
Because this study focuses on measurements with a clear differential flow signature, we allow an additional class of fits for which the alpha particle temperature has been fixed to the proton core temperature so long as the alphas are well separated from the proton beam.
To ensure that the magnetic field is suitably constant over the measurement time, we follow \citet{Kasper2002b} and we reject spectra for which the RMS fluctuation of the local magnetic field direction is larger than $20^o$.
In addition to the reported impact on alpha particle measurements, we find that excluding these spectra also improves the overall quality of reported proton beams.
To ensure that the beam is well constrained, we only include spectra for which the beam phase space density is larger than the core phase space density at the beam's bulk velocity, i.e.\ $f_{p_2}/f_{p_1}\left(\Bv_{p_2}\right) \ge 1$.
The vertical dashed lines in Figure\nobreakspace \ref {fig:thetaBn} indicate where this ratio is evaluated in each look direction.
The look directions that are most aligned with the magnetic field direction give the clearest view of the beam.

\section{Fast Wind Differential Flow} \label{sec:fsw}
Fig.\nobreakspace \ref {fig:dvca-fsw} shows the distributions of simultaneously-measured  differential flows in the fast wind ($v_{\text{sw}} \geq 400 \; \mathrm{km \, s^{-1}}$) under conditions where the  alphas and protons are both roughly collisionless ($10^{-2} \lesssim A_c \lesssim 10^{-1}$).\footnote{See Section \ref{sec:Ac} for a discussion of collisional age.}
The dashed lines are alpha-proton core differential flow ($\dv{{\alpha,p_1}}/C_A$) and the solid lines are proton beam-core differential flow ($\dv{{p_2,p_1}}/C_A$).
Here, we normalize to the ideal MHD Alfv\'en speed following Eq.\nobreakspace \textup {(\ref {eq:Ca})} and consider only the proton beam and core densities.\footnote{See Section \ref{sec:Ca} for a discussion of the Alfv\'en speed.}
The gray lines are histograms of all data.
In order to extract representative values and spreads thereof, we fit the green regions corresponding to $30\%$ of the peak with a Gaussian.
In selecting this portion of the histogram, we implicitly exclude an allowed class of proton VDF fits in which dominant non-Maxwellian features appear as large tails or a halo in the proton distribution instead of a secondary peak or shoulder-like fit because the uncertainty on the drift velocity is large.
We leave these core-halo distributions for a later study.
For the $\alpha$-particle case, there is a distinct population with small drifts resulting from a combination of noise and poor quality fits. 
Requiring $\Delta v_{\alpha,p1}/C_A \geq 0.27$ addresses this issue.
The best fit Gaussians are shown in orange.
Similar to previous results (e.g.\ \citet{Kasper2008,*Kasper2017,*Marsch1982b,*Reisenfeld2001}), % Neugebauer1976 didn't normalize dv by ca.
$\dvca{{\alpha,p_1}} = 67\%\pm26\%$ and $\dvca{{p_2,p_1}} = 108\%\pm16\%$, where the ranges quoted are the one-sigma widths of these fits.
The widths of the Gaussians, which we will heretofore denote $\sigma_{\alpha, p_1}$ and $\sigma_{p_2,p_1}$, are attributed to a combination of (1) the range of measured solar wind conditions that support a non-zero differential flow and (2) applicable measurement errors.
In the following sections, we hypothesize and test some potential contributions to each.

\plotdvcaFSW

\section{Uncorrelated Fluctuations} \label{sec:fluctuations}
Differential flow is strongest in solar wind with large Alfv\'enic fluctuations and therefore thought to be a signature of local wave-particle interactions, e.g.\ cyclotron-resonance-induced phase space diffusion for the case of proton beaming \citep{Tu2004}.
If differential flow is in general a product of local wave-particle interactions, the difference in widths observed in the histograms of Fig.\nobreakspace \ref {fig:dvca-fsw} may follow from a resonance condition or aspect of the wave-particle coupling that depends on ion species characteristics, such as charge-to-mass ratio.
To test this, we compare the magnitudes of correlated $\alpha$ and $p_2$ streaming fluctuations about their mean.

Figure\nobreakspace \ref {fig:ddv-ddv} is a 2D histogram of proton beam differential flow fluctuations ($\delta \dv{{p_2,p_1}}$) and alpha differential flow fluctuations ($\delta \dv{{\alpha,p_1}}$), each about their mean.
Comparing fluctuations in $\dv{}$ removes other sources of variation in the magnitude of $\dv{}$, such as large scale variations in the Alfv\'en speed or the bulk speed of the solar wind.
Fluctuations are calculated by subtracting a running 14 minute mean from each $\dv{}$ time series, and requiring spectra for $\sim 50\%$ of the time period.
Because the fitting algorithms returns the parallel component of the beam differential flow, comparing any other component would incorporate additional information about the magnetic field.
An ellipse is fit to the 2D histogram and contours of the fit are shown.
The insert gives the function and fit parameters.
The ellipse is a circle centered at the origin, indicating that the variations in $\dv{{\alpha,p_1}}$ and $\dv{{p_2,p_1}}$ are uncorrelated on these scales.
We conclude that the difference in $\dv{}$ distribution widths, i.e. $\sigma_{\alpha,p_1} \ne \sigma_{p_2,p_1}$, described in the previous section is not due to any species-specific difference in response to large scale, local fluctuations.
We repeated this calculation for running means calculated over various time intervals ranging from 5 minutes to more than 20 minutes and multiple requirements for the minimum number of spectra per window.
The result is not sensitive to either parameter.

\plotDeltaDvDeltaDv

\section{Trends with Collisional Age} \label{sec:Ac}

In a hot and tenuous plasma -- even in the absence of classical hard collisions -- the cumulative effect of small angle Coulomb collisions acts like a simple drag force that gradually slows differentially flowing particles \Citep{Spitzer1962}.
\Citet{Tracy2016a} showed that collisions with bulk protons are the dominant source of Coulomb drag on all other ions in the solar wind.
\Citet{Kasper2008,Kasper2017} have demonstrated that $\dvca{{\alpha,p_1}}$ is a strong, exponentially decaying function of the Coulomb collisional age, the ratio of the local collision rate to the local expansion rate.

The differential equation describing Coulomb drag is $\der{\dv{}}{t} = - \nu_c \dv{}$, where $\nu_c$ is the effective collision rate. In integral form, this becomes $\dv{} = \Delta v_0 \exp\!\left[- \int_0^{t_0} \nu_c \dd{t}\right]$. Under the highly-simplified assumption that $\nu_c$ and the solar wind speed ($v_{\text{sw}}$) are constant over the \replaced{course of propagation}{propagation distance $r$}, the integral is commonly estimated as $\int_0^{t_0} \nu_c\dd{t} = \nu_c r/v_{\text{sw}}$.
We follow \citet{Kasper2008} and refer to this empirical proxy for the total number of collisions experienced over the expansion history as the collisional age ($A_c$) of the solar wind.
\begin{equation}\label{eq:Ac}
A_c = \nu_c \times \frac{r}{v_\mathrm{sw}}
\end{equation}
\Citet{Kasper2017} refer to the same quantity as the Coulomb Number ($N_c$).
\added{\citet{Chhiber2016} provide a detailed comparison of this empirical proxy to simulations.}
As we show below, the exponential decay of $\dv{}$ with collisional age implies that $\dvca{}$ histogram widths $\sigma_{\alpha,p_1}$ and $\sigma_{p_2,p_1}$ is highly sensitive to the range of $A_c$ in the sample.

Based on the work of \citet{Tracy2016a}, we neglect collisions amongst the minor populations themselves and only consider collisions of $\alpha$ or $p_2$ ions with proton core ions ($p_1$).
Based on the work of \citet{Kasper2008,Kasper2017}, we limit our analysis of the collisional age dependence to collisionless and weakly collisional regimes that constitute the range $10^{-2} \lesssim A_c \lesssim 10^{-1}$. This is the range in which $\dvca{{\alpha,p_1}}$ is empirically non-zero.

Because the proton beam can have a non-negligible density in comparison to the proton core, we calculate the collision frequency between two species following \citet[Eq. (23)]{Hernandez1985} in a self-consistent manner by integrating over test and field particles from both components.
Our treatment of the Coulomb logarithm follows \citet[Eq. (18)]{Fundamenski2007}.
We assume that $r$ is the distance traveled from a solar source surface to the spacecraft's radial location, $\approx$ 1 AU, and we take the solar wind velocity to be $v_{\text{sw}} \approx v_{p_1}$. 

Measurements of $\dvca{{\alpha,p_1}}$ and $\dvca{{p_2,p_1}}$ are binned by collisional age and histogrammed in Figure\nobreakspace \ref {fig:dvca-Ac} across the aforementioned range.
Each column has been normalized by its maximum value in order to emphasize the trends with $A_c$.
Only bins with at least $30\%$ of the column maximum are shown. To characterize the collisionally ``youngest'' solar wind spectra that have been measured, we define a sufficiently large and statistically significant subset that reflects the limiting behavior.
We have chosen this ``youngest'' range to be ($10^{-2} \le A_c \le 1.2 \times 10^{-2}$).
The rightmost limit of this subset is marked with a blue line on the figure.

In the case of $\alpha$ particles, the decrease from the mean value in the reference or youngest region of $\dvca{{\alpha,p_1}} \sim 0.8$ down to $\dvca{{\alpha,p_1}}\sim 0.4$ over the range shown would appear to account for a significant fraction of $\sigma_{\alpha,p_1}$, up to a $\sim$ 40\% spread.
In contrast, the proton analogue exhibits a far weaker apparent decay with increasing collisions,showing a decrease of at most approximately one-tenth the slope of the alpha particle trend. In other words, $\dvca{{p_2,p_1}}$ is nearly independent of the collisional age.

\plotdvcaActwoD

We would also like to derive the general and limiting cases for the differential flow speed ratios $\dv{{p2,p1}}/\dv{{\alpha,p1}}$ in spectra where the two are observed simultaneously.
In Fig.\nobreakspace \ref {fig:dvrat}, we compare $\dv{{\alpha,p_1}}$ to $\dv{{p_2,p_1}}$ directly in the full low-collision regime and in the very young reference regime.
The ratios $\dv{{\alpha,p_1}}/\dv{{p_2,p_1}}$ are histogrammed, with the dashed line indicating the full low-collision sample $10^{-2} \le A_c \le 10^{-1}$ and the solid line indicating the reference or youngest subsample ($10^{-2} \le A_c \le 1.2 \times 10^{-2}$).
The selection of data that contributes to Fig.\nobreakspace \ref {fig:dvrat} is slightly different and more restrictive than in the previous section, because here we require that both the alpha-core and proton beam-core collision rates simultaneously fall in the target range.

As before, we characterize these distributions in Fig.\nobreakspace \ref {fig:dvrat} in a manner insensitive to the tails by fitting a Gaussian to bins with a count of at least $30\%$ of the most populated bin.
Similar to Fig.\nobreakspace \ref {fig:dvca-fsw}, all binned data are shown in gray; the regions fit are green; and the fits are orange.
The text inserts give the functional form and fit parameters up to the fit uncertainty.
As there are fewer counts in the youngest $A_c$ range, the histograms have been normalized by their maximum values in order to emphasize the difference in the respective means ($\mu$) and widths ($\sigma$) of the distributions.

\plotdvrat

Over the low-collision range, $\dv{{p_2,p_1}}$ is approximately $1.6\times$ faster than $\dv{{\alpha,p_1}}$.
Over the youngest range, that reduces to $1.4\times$.
The width or characteristic spread in $\dv{{\alpha,p_1}}/\dv{{p_2,p_1}}$ is $1.37\times$ larger over the broader, low-collision range than the youngest range.
Having demonstrated that $\dv{{\alpha,p_1}}$ and $\dv{{p_2,p_1}}$ are uncorrelated in these ranges and that the mean value of $\dvca{{\alpha,p_1}}$ changes by about $0.4$ over the full range, we attribute most of the spread in the ratio $\dv{{\alpha,p_1}}/\dv{{p_2,p_1}}$ to the observed decay of $\dv{{\alpha,p_1}}$ with increasing Coulomb collisions.

\section{Corrections to the Alfv\'en Speed} \label{sec:Ca}
Alfv\'en waves are parallel propagating, transverse, non-compressive fluctuations in MHD plasmas.\citep{Alfven1942}
Under ideal MHD and considering only a single, simple fluid, the phase speed of these waves (the Alfv\'en speed) is given by the ratio of the magnetic field magnitude ($B$) to the square root of the mass density ($\rho$):
\begin{equation}\label{eq:Ca}
C_A = \frac{B}{\sqrt{\mu_0 \rho}}.
\end{equation}

\Citet{Barnes1971} derived an approximation to the phase speed of the Alfv\'en wave under anisotropic MHD that accounts for pressure anisotropy and differential flow of multiple ion species:
\begin{equation}\label{eq:CaAni}
C_{A}^{\text{Ani}} = C_A \left[1 + \frac{\mu_0}{B^2}\left(p_\perp - p_\parallel\right) - \frac{\mu_0}{B^2}p_{\tilde{v}}\right]^{1/2}.
\end{equation}.

\noindent
Here, $C_A$ is the ideal MHD Alfv\'en speed from Eq.\nobreakspace \textup {(\ref {eq:Ca})}.
The second term in the brackets gives the correction due to the thermal anisotropy of the plasma. Total thermal pressure perpendicular and parallel to the local magnetic field are $p_i = \sum_s n_s k_b T_{s,i} = \frac{\rho_{p_1}}{2}\sum_s\frac{\rho_s}{\rho_{p_1}}w_{s;i}^2$ for components $i = \perp,\parallel$.
The third term in the brackets gives the correction due to the dynamic pressure from differential streaming in the plasma frame which is $p_{\tilde{v}} = \sum_s \rho_s \left(\Bv_s - \Bu\right)^2 = \rho_{p_1} \sum_s \frac{\rho_s}{\rho_{p_1}} \left(\Bv_s - \Bu\right)^2$.
Here, $\Bu$ is the plasma's center-of-mass velocity; a given species' mass density is $\rho_s$; and its velocity is $\Bv_s$.
All species $s$ are summed over.
Pressure terms have been written in terms of mass density ratios to emphasize the significance of correction factors discussed in the following paragraphs and cataloged in Table\nobreakspace \ref {tbl:mu-sigma}.
When the plasma is isotropic and there is either vanishingly slow differential flow or a vanishingly small differentially flowing population, the term in brackets is equal to unity and Eq.\nobreakspace \textup {(\ref {eq:CaAni})} reduces to Eq.\nobreakspace \textup {(\ref {eq:Ca})}. 

This anisotropic, multi-component formalism of \citet{Barnes1971} ought to be a more appropriate and higher fidelity description of the solar wind plasma than the commonly-evoked ideal single-fluid approximation.
Nevertheless, it is instructive to give a rough illustration of the magnitude of each correction term under typical conditions.
We note first that the proton core in the solar wind is often anisotropic, with core pressure ratios falling primarily in the range $0.1 \lesssim p_{\perp}/p_{\parallel} \lesssim 10$.
The absolute correction to the Alfv\'en speed, via the second bracketed term in Eq.\nobreakspace \textup {(\ref {eq:CaAni})}, that follows from this anisotropy alone is $\sim$6\%-7\% for the median case and can be as high as $\sim$50\%.
With regards to the third bracketed term, we note that a typical proton beam carrying $10\%$ of the total protons at a speed of roughly $C_A$ relative to the core would carry a $\sim$5\% self-consistent correction to the Alfv\'en speed, owing to proton beam-core dynamic pressure.

Our goal in this section is to relax the ideal MHD approximation by considering these next-order approximations for the speed of the predominant parallel-propagating wave in the solar wind.
We explore whether the spreads in normalized differential flow, i.e.\ the widths of the 1D distributions of $\dvca{}$, are further minimized when the contributions of anisotropic and dynamic pressure are considered.
In order to disentangle this element from the Coulomb collision effect described in the previous section, we limit our analysis in this section to the ``youngest'' plasma, i.e.\ measurements drawn from the youngest-measured reference regime to the left of the blue line in Fig.\nobreakspace \ref {fig:dvca-Ac}.

Figure\nobreakspace \ref {fig:ex-dvca-yAc-fits} plots distributions and fits in the now-familiar style, together with the fit residuals, for one possible renormalization of $\dvca{{\alpha,p_1}}$ and $\dvca{{p_2,p_1}}$.
The color selection for the various components in the top panel follows the convention from the previous figures and again only bins with counts at least $30\%$ of the maximum are used in the fit.
Residuals are shown for the bins in the fit, and the fit parameters are shown in the inserts.
The amplitudes $A$ are omitted because they are of no consequence.
In this particular case, the $\alpha$ and $p_2$ differential flow are normalized by the Alfv\'en speeds with proton core pressure anisotropy taken into account.
For reasons discussed below, the normalization in the proton beam-core example (Right) also accounts for the beam contribution to the proton mass density.

\plotexOneD

We consider a family of similar approximations to the Alfv\'en speed, each accounting for corrections associated with the measured anisotropies and multiple component terms in Eq.\nobreakspace \textup {(\ref {eq:CaAni})}.
As these contributions rely on higher-order moments of the spectrum fit\footnote{See Section \ref{sec:data}.}, they can carry relatively large uncertainties.
If the uncertainties are significant in the aggregate, they are expected to contribute to broadening of the $\dvca{}$ distributions.
However, terms that are well-measured in the aggregate, will improve the precision of the Alfv\'en speed when accounted for and thus reduce the width of $\dvca{}$ if the true differential flows are Alfv\'enic in nature.
In the following, we examine all possible combinations in order to ascertain whether a well-measured high order correction exists that further minimizes the width of the normalized differential flow distributions. 

Table\nobreakspace \ref {tbl:mu-sigma} contains fit parameters for each 1D distribution of $\dvca{}$, for both the alpha-proton and proton beam-core differential flows, using the various formulations of the Alfv\'en speed.
Overall, we find that the widths of both $\dvca{}$ distributions increase substantially when the dynamic pressure term is included, indicating that either (1) the differential flows are \emph{less} strongly correlated with generalized Alfv\'en speed, or (2) that the additional measurement uncertainty introduced along with a given term is in the aggregate comparable to the correction itself.
%\authorcomment1{Pending what the referee says, Mike and I have some notes we can use to tighten up the previous paragraph.}

\input{mu_sigma_table.tex}

\noindent
However, when only the proton core temperature anisotropy correction is factored in, the distribution width is indeed reduced relative to the isotropic case.
Because the core anisotropy correction term in Eq.\nobreakspace \textup {(\ref {eq:CaAni})} is usually (but not always) positive, it tends to increase the Alfv\'en speed estimate relative to the ideal MHD approximation.
Thus, the corrected mean values $\dvca{}$ are generally lower.
Figure\nobreakspace \ref {fig:mu-sigma} is a plot of the $\mathrm{width}$ vs.\ $\mathrm{mean}$ for select 1D fits that were performed in the style of Figure\nobreakspace \ref {fig:ex-dvca-yAc-fits}, illustrating these observations. 
In the cases shown, each Alfv\'en speed includes both proton densities. The cases accounting for proton core pressure anisotropy correction factor $(p_\perp - p_\parallel)$ are indicated with the square. Cases that additionally account for the proton core dynamic pressure correction factor ($p_\perp - p_\parallel - p_{\tilde{v}}$) are indicated by stars.

\plotmusigma

\section{Trends in $A_c$} \label{sec:trends-Ac}
Using the Alfv\'en speed approximation that minimizes the spread in normalized differential flow for alphas and beams, we examine the behavior of $\dvca{}$ as a function of $A_c$ and in the asymptotic limit of zero collisions.
We applied the same methodology used to examine 1D distributions in the youngest $A_c$ data to binned $\alpha,p_1$ and $p_2,p_1$ differential flow spanning the low-collision range.
Figure\nobreakspace \ref {fig:trends-Ac} plots these trends.
Alpha particles are shown in blue and proton beams in yellow.
Mean values to 1D fits are indicated as pluses and the 1D widths are given as error bars.
Fits to each trend are given as black dotted lines.

Four clear features are apparent pertaining to the mean values of both normalized differential flows and to their collisional trends.
First, if we consider the asymptotic limit of zero Coulomb collisions and we account for the widths reported in Table\nobreakspace \ref {tbl:mu-sigma}, the alpha particles differentially stream at $67\%$ of the local Alfv\'en speed and the proton beams stream at approximately the Alfv\'en speed.
Second, that the fit constant $c$ governing $\alpha,p_1$ decay is greater than 1 indicates that our collisional age calculation over-simplifies our $A_c$ by either under-estimating $r$, under-estimating $\nu_c$, over-estimating $v_\mathrm{sw}$, or some combination of these.
\citet{Kasper2017} examined detailed scalings and more accurate versions of $A_c$ that may correct for some of these issues and can be a subject for future study.
Third, even using the formulation of the Alfv\'en speed that yields the highest precision, the spread in alpha particle differential flow due to the change in mean value over the collisionless range is still $\sim 0.3$, which is the largest single contribution to the spread in $\dvca{}$.
Fourth, in the asymptotic absence of collisions, the proton beams differentially flow at very nearly ($105\%$ of) the Alfv\'en speed.
Given the widths of the error bars in Fig.\nobreakspace \ref {fig:trends-Ac}, the difference between the youngest resolved $\dv{{p_2,p_1}}$ and the asymptotic value could be due to the spread in our measurements.

\plotActrends

\input{new_discussion}

\section{Conclusions}
\label{sec:cncl}

In fast ($> 400 \; \mathrm{km \; s^{-1}}$) and collisionless ($A_c \le 10^{-1}$) solar wind, $\alpha,p1$ differential flow is approximately $62\%$ as fast as $p2,p1$ differential flow when measured by the \textit{Wind} spacecraft's Faraday cups.
The spread in $\alpha,p1$ differential flow is approximately $1.7\times$ larger than $p2,p1$ differential flow.
We ruled out large-scale, in-phase wave-particle interactions by examining the correlation between fluctuations in both species parallel differential flows over multiple time scales ranging from 5 minutes to more than 20 minutes.
Minimizing the spread in normalized differential flow due to the method used to approximate the Alfv\'en speed, we found that the difference in $\dvca{}$ width for both species is predominantly due to the decay of $\dvca{{\alpha,p1}}$ with increasing Coulomb collisions.
At the youngest resolved collisional age, when the impact of Coulomb collisions has been minimized, we find that proton core pressure anisotropy has the largest impact on minimizing the spread in normalized differential flow and that the increase in spread when including dynamic pressure in the anisotropic Alfv\'en speed is beyond what would be expected from random fluctuations.
In the asymptotic absence of Coulomb collisions, $\alpha$-particles differentially flow at approximately $67\%$ of the local Alfv\'en speed and proton beams differentially flow at approximately $105\%$ of it.
This upper limit on $\dvca{{\alpha,p1}}$ is close to the upper limit found by \citet{Maneva2014} and worth further investigation.
We also found that, unlike the known \citep{Neugebauer1976,Kasper2008,Kasper2017} $\alpha,p1$ decay with $A_c$, proton beam differential flow minimally decays and is approximately constant with collisional age.

Given the results of \citet{Tracy2016a} showing that solar wind ions collisionally couple most dominantly to protons, it is unsurprising that the widths of both $\dvca{{\alpha,p1}}$ and $\dvca{{p2,p1}}$ are smallest when the Alfv\'en speed accounts for the proton core.
That the proton core temperature anisotropy is also significant supports the conclusion of \citet{Chen2013} that solar wind helicities are closer to unity when normalzing by the anisotropic Alv\'en speed.
That the beam differential flow width is smaller when it is normalized by an Alfv\'en speed including the beam density may indicate some coupling between the beams and local Alfv\'en waves, as predicted by \citet{Voitenko2015}.
That the dynamic pressure term causes a larger spread in both species normalized differential flow is either a result of measurement uncertainty or some underlying physical mechanism that is beyond the scope of this paper to test.

\acknowledgments

The authors thank K.\ G.\ Klein, D.\ Verscharen, and P.\ Whittlesey for useful discussions.
B.\ L.\ Alterman and J.\ C.\ Kasper are supported by NASA grant NNX14AR78G.
M.\ L.\ Stevens is supported by NASA grant NNX14AT26G.
Both grants support \textit{Wind} operations and data analysis.


\software{pandas \citep{{Mckinney2010}},
          Python \citep{Millman2011,*Oliphant2007}, 
          SciPy \citep{Jones2001},
          NumPy \citep{VanderWalt2011},
          Matplotlib \citep{Hunter2007},
          IPython \citep{Perez2007},
          jupyter \citep{Kluyver2016}
          }




%\clearpage
\bibliography{Alterman2018}


\listofchanges

\end{document}

%% file: preamble.tex
\newcommand{\Bb}{\ensuremath{\bm{b}}}

\newcommand{\Bu}{\ensuremath{\bm{u}}}
\newcommand{\Bv}{\ensuremath{\bm{v}}}

\newcommand{\BB}{\ensuremath{\bm{B}}}

\newcommand{\uv}[1]{\ensuremath{\hat{\bm{#1}}}}
\newcommand{\bhat}{\uv{\Bb}}

\newcommand{\der}[2]{\frac{\mathrm{d} #1}{\mathrm{d} #2}} 
\newcommand{\dd}[2][]{\, \mathrm{d}{ #1 } #2}

\newcommand{\dv}[1]{\ensuremath{\Delta v_{#1}}}
\newcommand{\dvca}[1]{\ensuremath{\dv{#1}/C_A}}

%% file: figs.tex
\newcommand{\plotThetaBn}{
\begin{figure}
\includegraphics[width=\columnwidth]{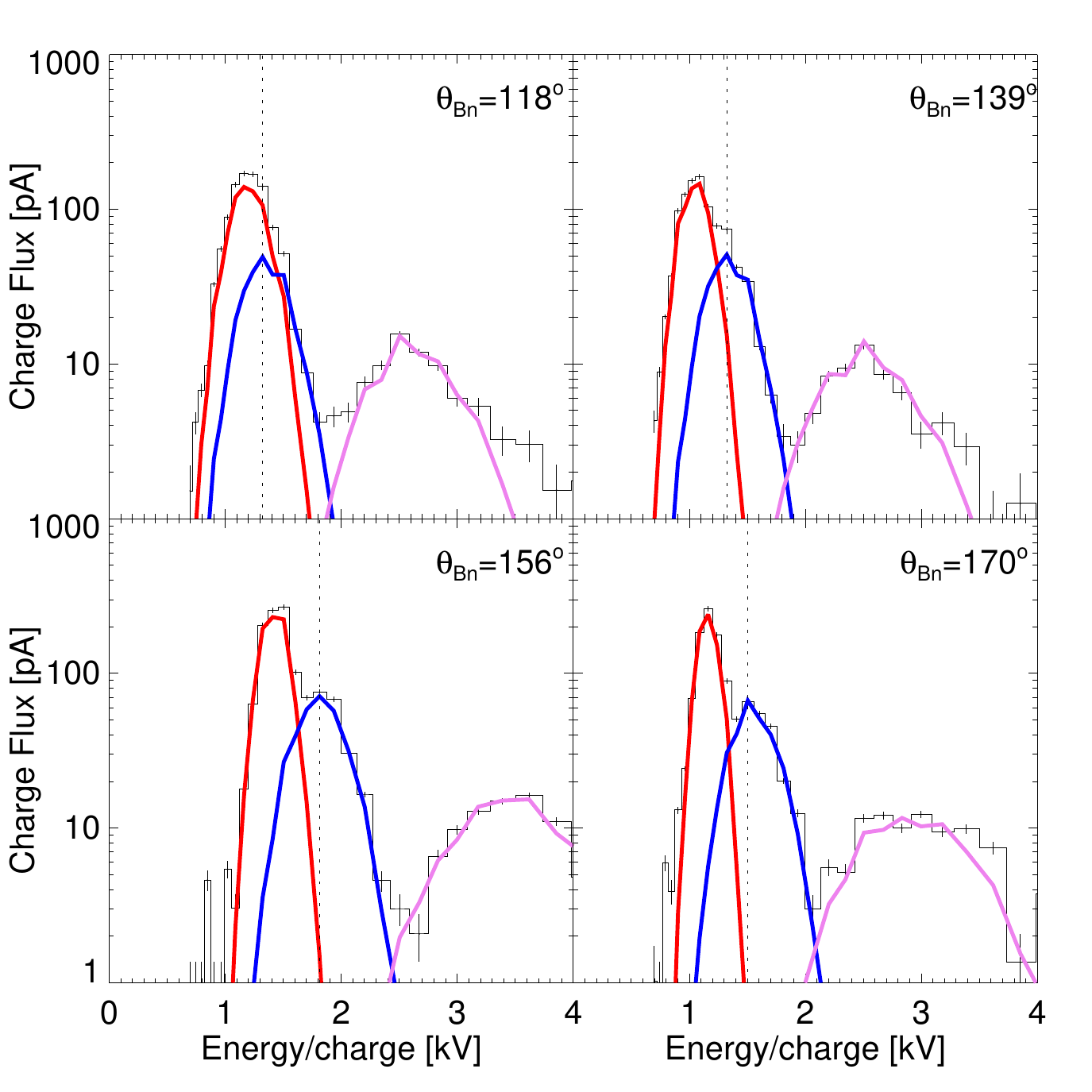}
\caption{Fits from four example look directions from the \textit{Wind} Faraday cups using a new data processing algorithm. Three ion populations are shown: $\alpha$ (purple), $p_1$ (red), and $p_2$ (blue). The angle of a given look direction with respect to the average magnetic field throughout the spectrum is indicated in the top right of each panel. Errors for each Energy/charge bin are vertical dashed lines. 
    \label{fig:thetaBn}}
\end{figure}
}

\newcommand{\plotVDFex}{
\begin{figure}
\includegraphics[width=\columnwidth]{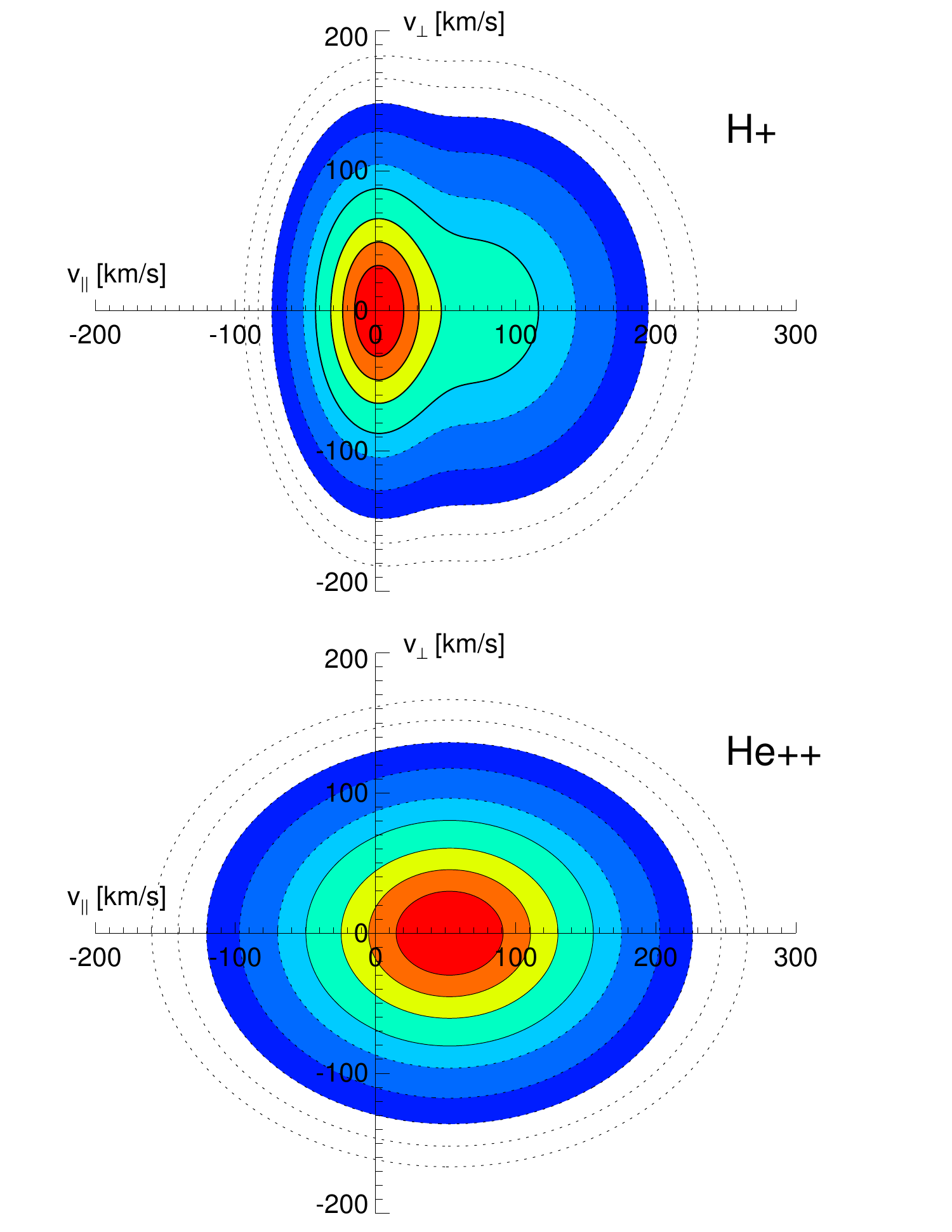}
\caption{VDFs corresponding to the spectrum shown in Fig.~\ref{fig:thetaBn}. The joint proton VDF is shown on (top) and the $\alpha$ particle VDF is shown on (bottom). The proton beam can be identified by the secondary shoulder with a large $v_\parallel$ in (top) plot. Contours follow \citet{Marsch1982c}. In decreasing order, solid lines are 0.8, 0.6, 0.4, 0.2 and dashed lines are 0.1, 0.032, 0.01, 0.0031, 0.001 of the maximum phase space density.
    \label{fig:exVDFs}}
\end{figure}
}

\newcommand{\plotdvcaFSW}{
\begin{figure}
\hspace*{-1em}\includegraphics[width=\columnwidth]{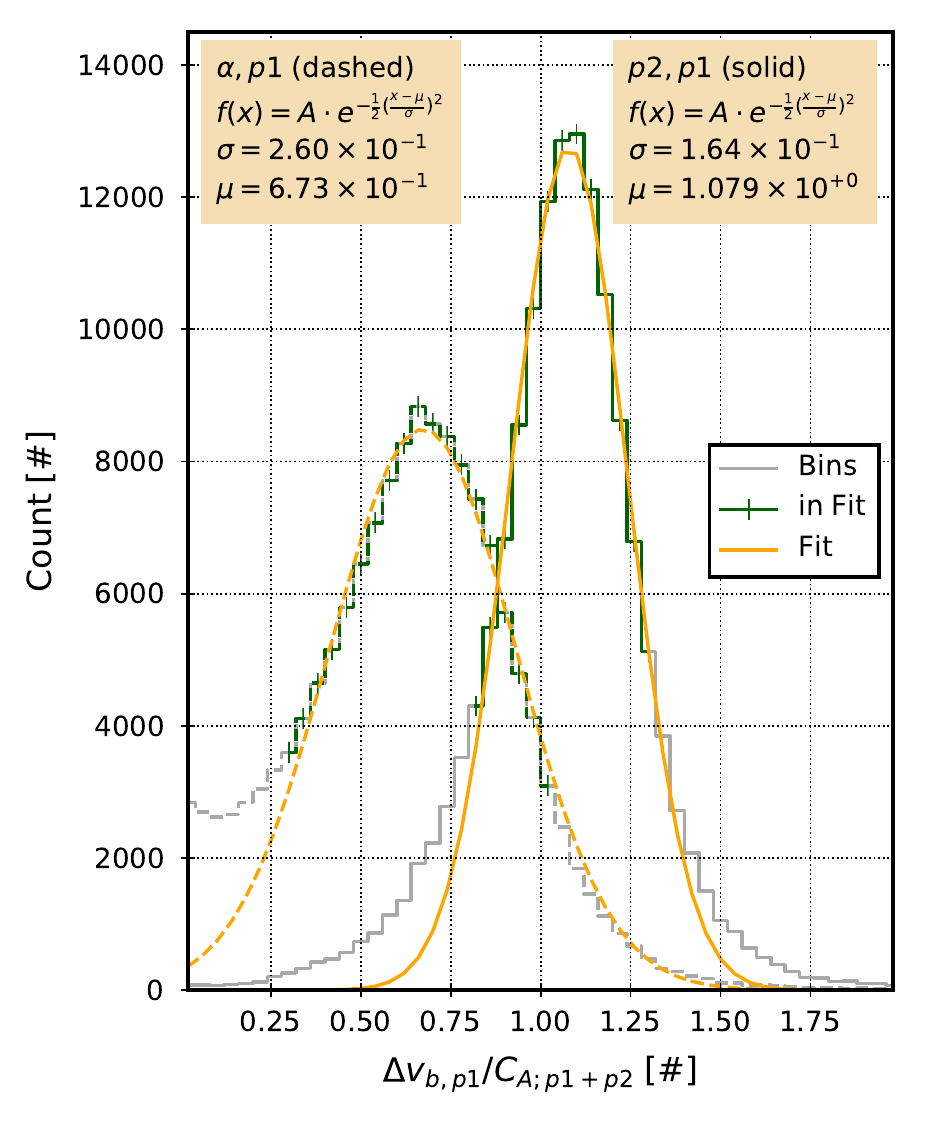}
\caption{Normalized Alpha particle ($\alpha,p_1$) and proton beam ($p_2,p_1$) differential flow in collisionless, fast solar wind. Both differential flows are normalized by an Alfv\'en speed approximation from Eq.~\ref{eq:Ca} using both proton densities. Bins within $30\%$ of the maximum are selected for fitting to exclude core-halo distributions.
    \label{fig:dvca-fsw}}
\end{figure}
}

\newcommand{\plotDeltaDvDeltaDv}{
\begin{figure}
\hspace*{-1em}\includegraphics[width=\columnwidth]{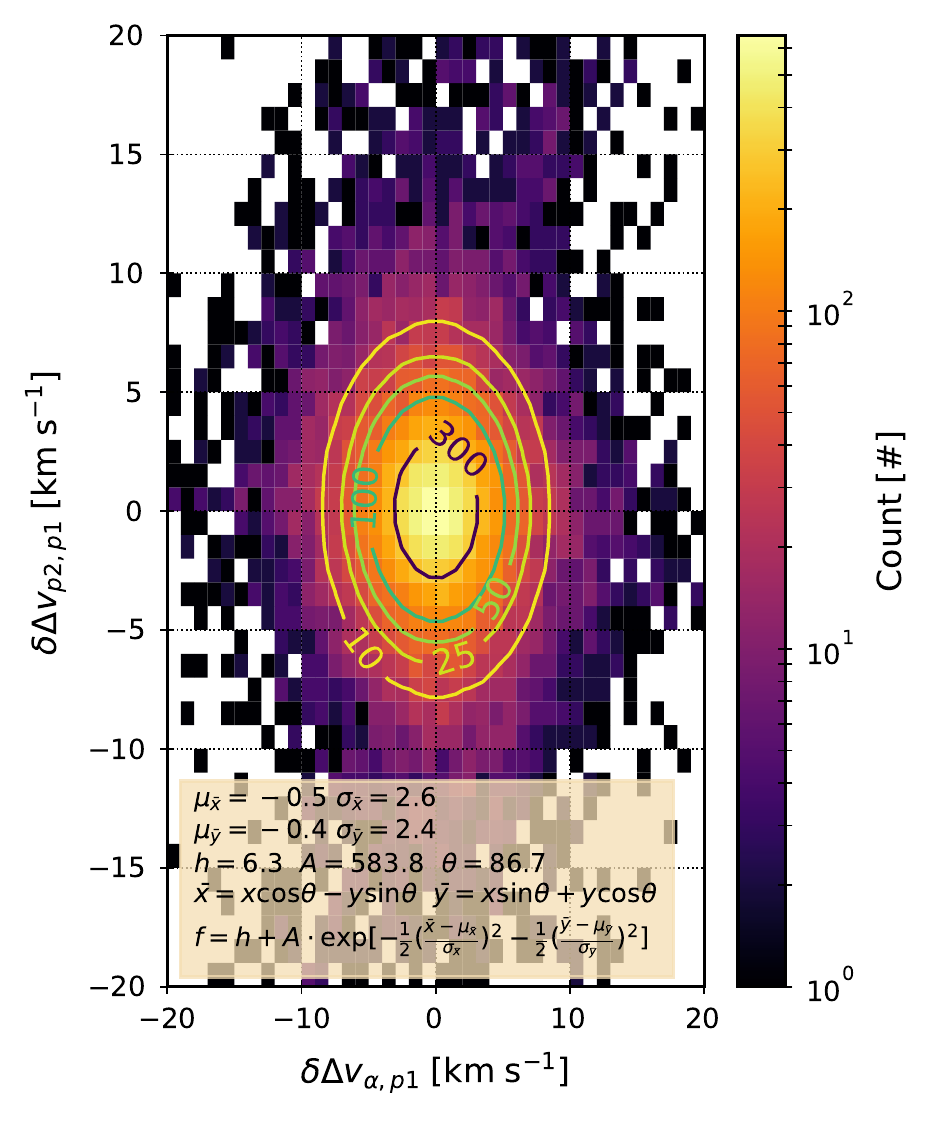}
\caption{A 2D histogram showing uncorrelated differential flow fluctuations ($\delta \dv{}$) for $\dv{{\alpha,p_1}}$ and $\dv{{p_2,p_1}}$. That the fit is a circle centered on the origin indicates that the fluctuations are uncorrelated.
    \label{fig:ddv-ddv}}
\end{figure}
}

\newcommand{\plotdvcaActwoD}{
\begin{figure}
\hspace*{-1em}\includegraphics[width=1.05\columnwidth]{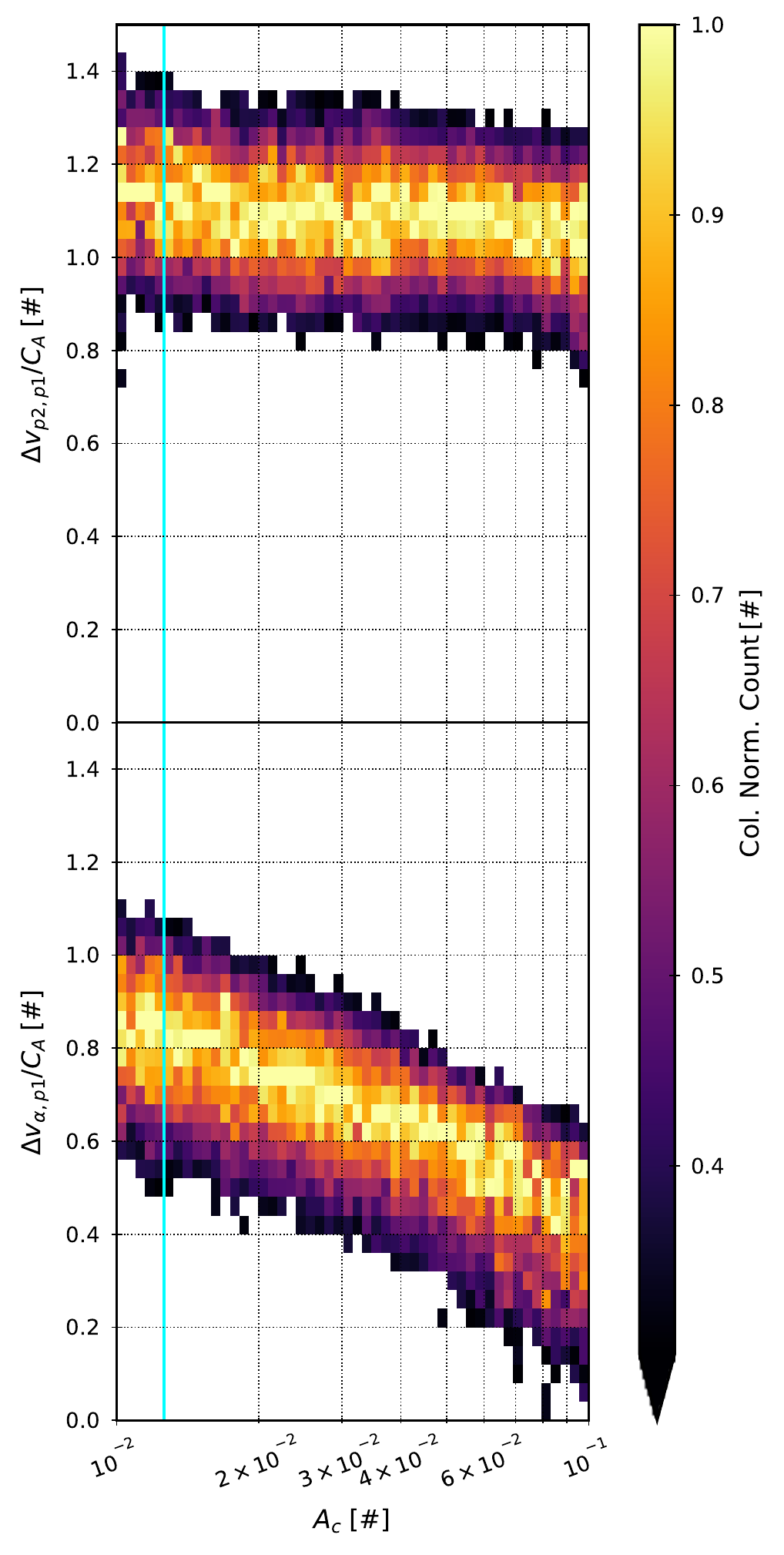}
\caption{2D histograms of $\alpha$ particle and $p_2$ Alfv\'en speed normalized differential flow each as a function of its collisional age. Only bins with at least $30\%$ of the a column maximum are shown. Measurements with a collisional age $A_c \lesssim1.2\times10^{-2}$ is indicated to the left of the blue line.
    \label{fig:dvca-Ac}}
\end{figure}
}

\newcommand{\plotdvrat}{
\begin{figure}
\hspace*{-1em}\includegraphics[width=\columnwidth]{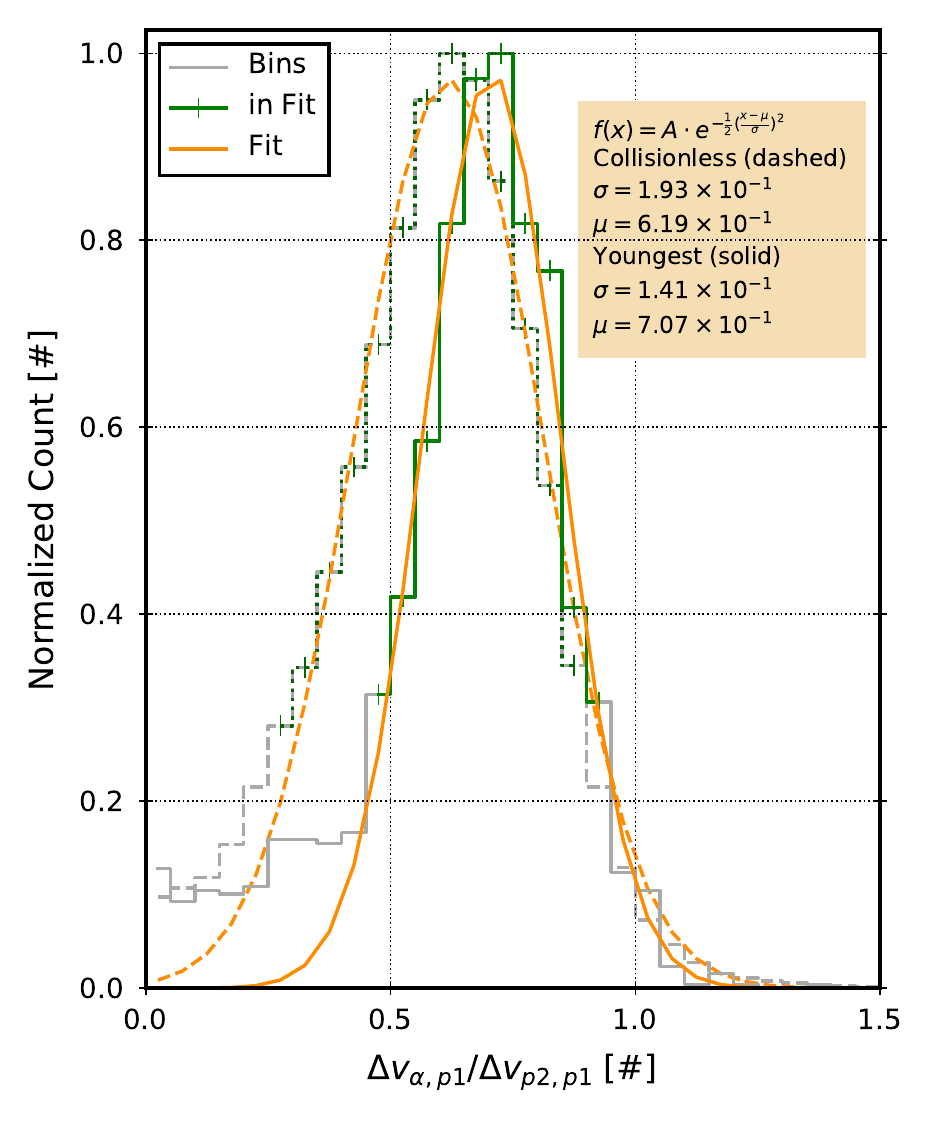}
\caption{The ratio of alpha particle to proton beam differential flow ($\dv{{\alpha,p_1}}/\dv{{p_2,p_1}}$) in collisionless ($10^{-2} \le A_c \le 10^{-1}$, dashed) and the youngest measured data ($10^{-2} \le A_c \le 1.2 \times 10^{-2}$, solid).
    \label{fig:dvrat}}
\end{figure}
}

\newcommand{\plotexOneD}{
\begin{figure}
\hspace*{-1em}\includegraphics[width=1.1\columnwidth]{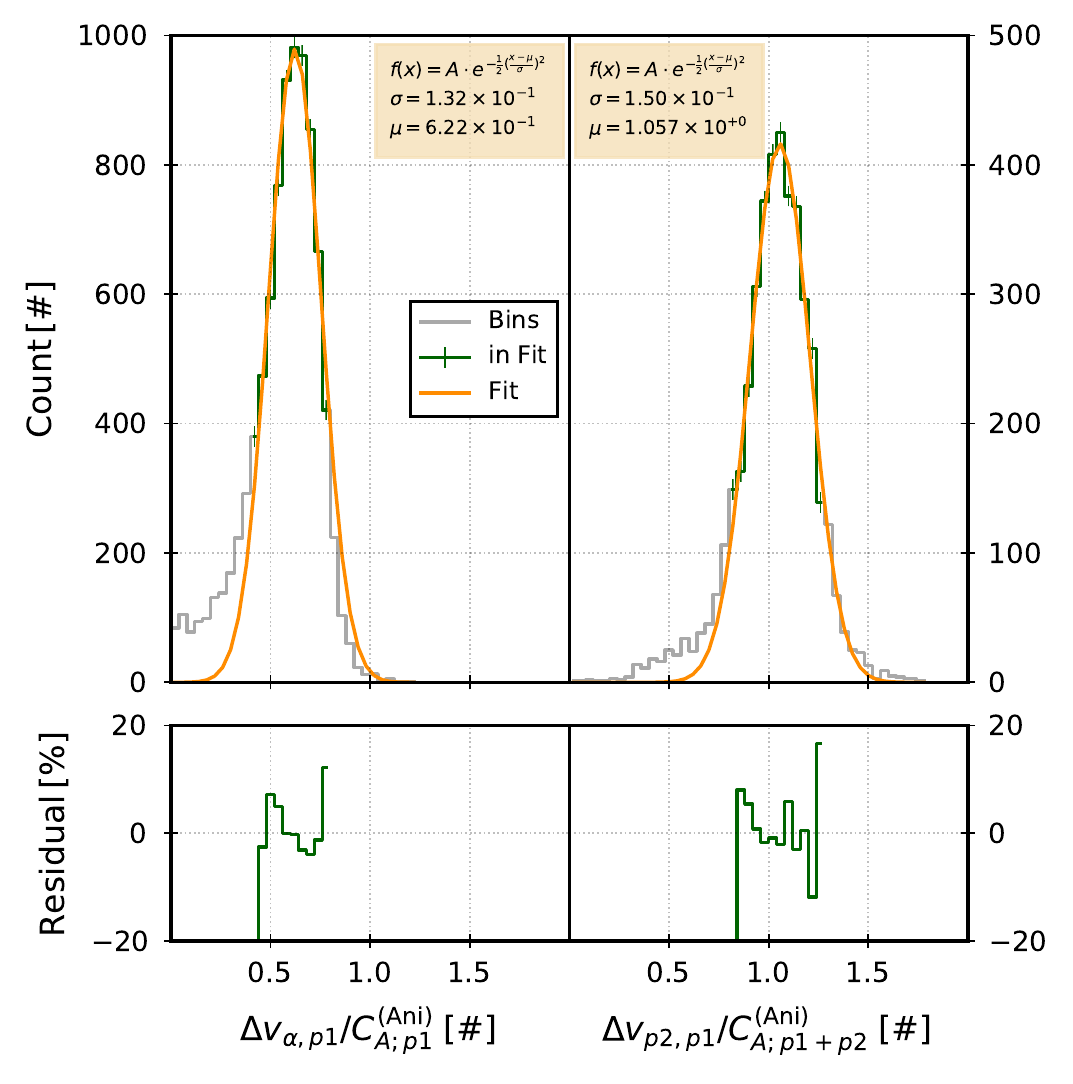}
\caption{Examples of the Gaussian fits to 1D distributions of $\alpha$ and $p_2$ normalized differential flow along with the associated residuals. As discussed in Section \ref{sec:Ca}, the Alv\'en speed normalizations shown minimize the width of these distributions.
    \label{fig:ex-dvca-yAc-fits}}
\end{figure}
}

\newcommand{\plotmusigma}{
\begin{figure}
\hspace*{-1em}\includegraphics[width=\columnwidth]{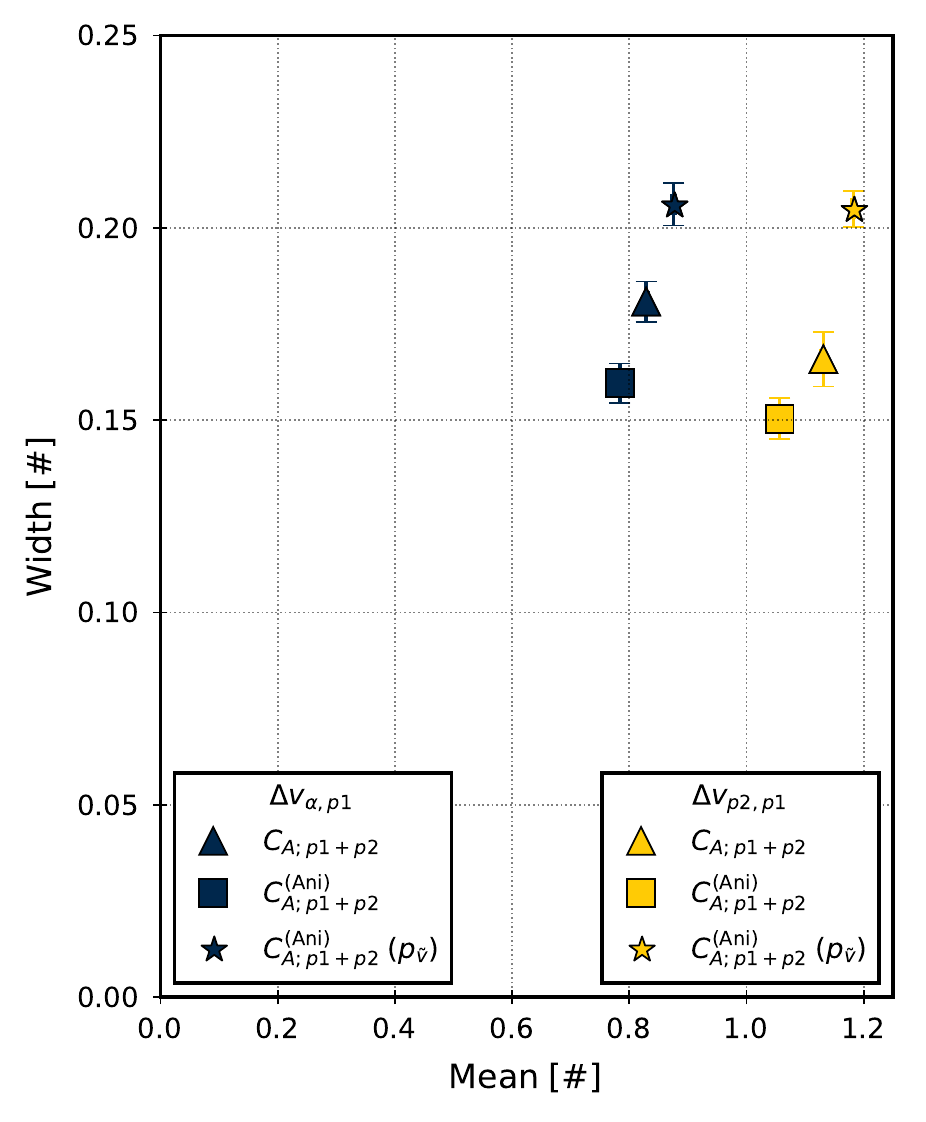}
\caption{Example $\alpha$-particle and $p_2$ normalized differential flow illustrating the impacts of various Alfv\'en speed approximations. In both cases shown, inclusion of the proton core anisotropy (Eq.~\ref{eq:CaAni}) reduces the width in comparison to the isotropic MHD Alf\'en speed (Eq.~\ref{eq:Ca}), while including the anisotropy and the dynamic pressure ($p_{\tilde{v}}$) increases it.
    \label{fig:mu-sigma}}
\end{figure}
}

\newcommand{\plotActrends}{
\begin{figure}
\hspace*{-1em}\includegraphics[width=\columnwidth]{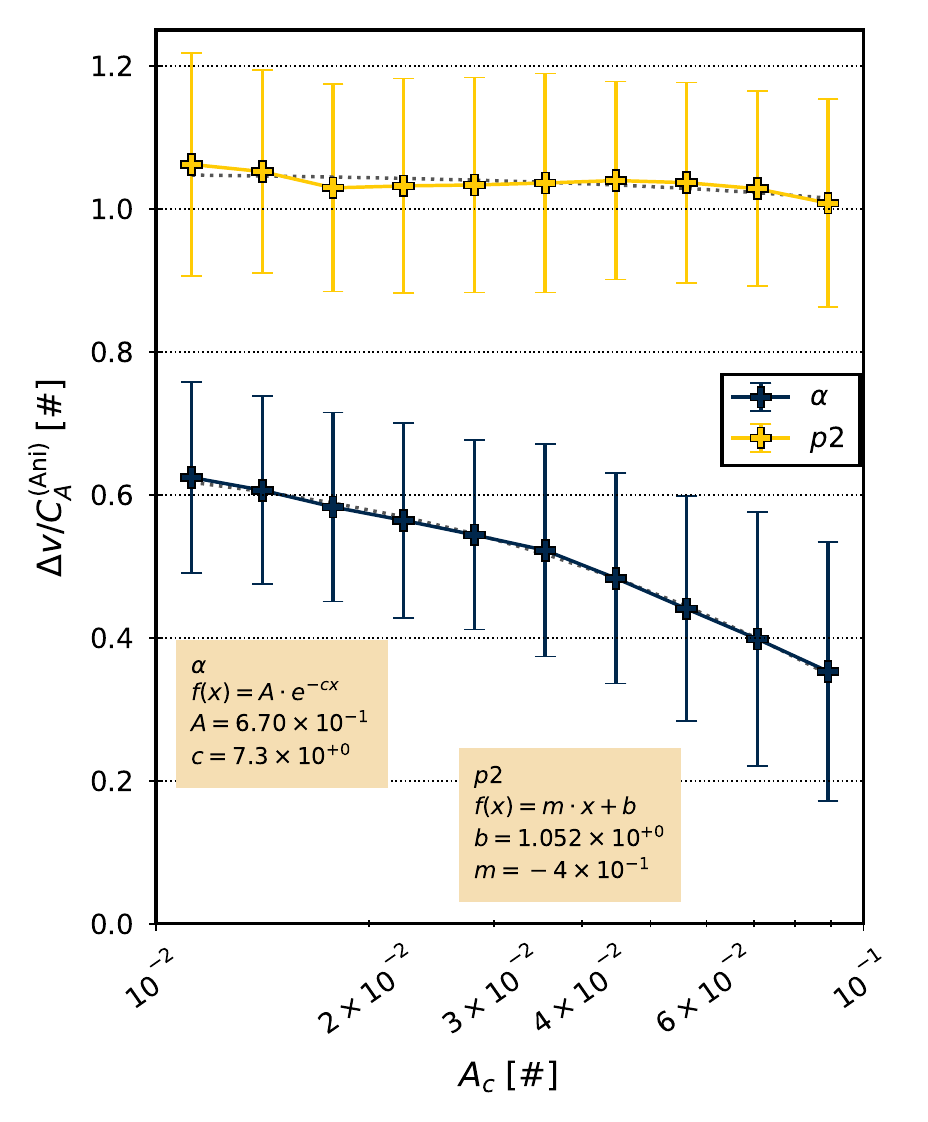}
\caption{Trends of 1D fits to $\dvca{{\alpha,p_1}}$ and $\dvca{{p_2,p_1}}$ as a function of $A_c$. Error bars are the $\mathrm{Widths}$ of the 1D fits. Each trend has been fit and the parameters are shown in the appropriate insert. While $\dv{{\alpha,p_1}}$ markedly decays with increasing $A_c$, $\dv{{p_2,p_1}}$ is relatively constant with $A_c$. To within the fit uncertainty, proton beams differentially stream at approximately the local Alfv\'en speed. 
    \label{fig:trends-Ac}}
\end{figure}
}

%% file: affiliations.tex
\correspondingauthor{B. L. Alterman}
\email{balterma@umich.edu}

\author[0000-0001-6673-3432]{B.\ L.\ Alterman}
\affiliation{University of Michigan \\
Department of Applied Physics\\
450 Church St.\\
Ann Arbor, MI 48109, USA}
\affiliation{University of Michigan \\
Department of Climate \& Space Sciences \& Engineering \\
2455 Hayward St. \\
Ann Arbor, MI 48109-2143, USA}

\author[0000-0002-7077-930X]{Justin C.\ Kasper}
\affiliation{University of Michigan \\
Department of Climate \& Space Sciences \& Engineering \\
2455 Hayward St. \\
Ann Arbor, MI 48109-2143, USA}
\affiliation{Smithsonian Astrophysical Observatory \\
Observatory Building E \\
60 Garden St.\\
Cambridge, MA 02138, USA}

\author[0000-0002-7728-0085]{Michael L.\ Stevens}
\affiliation{Smithsonian Astrophysical Observatory \\
Observatory Building E \\
60 Garden St.\\
Cambridge, MA 02138, USA}

\author{Andriy Koval}
\affiliation{Goddard Planetary Heliophysics Institute\\
University of Maryland Baltimore County\\
Baltimore, MD 21228, USA}
\affiliation{Heliospheric Physics Laboratory\\
NASA Goddard Space Flight Center \\
Greenbelt, MD 20771, USA}

%% file: mu_sigma_table.tex
\newcommand{\markprefered}[1]{\textcolor{mygreen}{\bm{#1}}}

\newcommand{\datauncertainties}{
$C^{( \mathrm{Ani})}_{A;\alpha+p1+p2} $            &                    $0.869 \pm 0.005$ &  $0.177 \pm 0.006$ &                $1.167 \pm 0.003$ &  $0.169 \pm 0.003$ \\
$C^{( \mathrm{Ani})}_{A;\alpha+p1+p2} \; (p_{\tilde{v}}) $ &                    $0.999 \pm 0.003$ &  $0.244 \pm 0.004$ &                $1.339 \pm 0.007$ &  $0.256 \pm 0.009$ \\
$C^{( \mathrm{Ani})}_{A;\alpha+p1} $               &                    $0.730 \pm 0.005$ &  $0.142 \pm 0.006$ &                $0.997 \pm 0.004$ &  $0.156 \pm 0.005$ \\
$C^{( \mathrm{Ani})}_{A;\alpha+p1} \; (p_{\tilde{v}}) $  &                    $0.761 \pm 0.004$ &  $0.164 \pm 0.005$ &                $1.048 \pm 0.003$ &  $0.172 \pm 0.004$ \\
$C^{( \mathrm{Ani})}_{A;p1+p2} $                   &                    $0.784 \pm 0.004$ &  $0.160 \pm 0.005$ &                $1.057 \pm 0.004$ &  $0.150 \pm 0.005$ \\
$C^{( \mathrm{Ani})}_{A;p1+p2} \; (p_{\tilde{v}}) $  &                    $0.876 \pm 0.005$ &  $0.206 \pm 0.005$ &                \markprefered{$1.182 \pm 0.004$} &  \markprefered{$0.205 \pm 0.005$} \\
$C^{( \mathrm{Ani})}_{A;p1} $                      &                    \markprefered{$0.622 \pm 0.004$} &  \markprefered{$0.132 \pm 0.005$} &                $0.874 \pm 0.003$ &  $0.164 \pm 0.004$ \\
$C_{A;\alpha+p1+p2}$                               &                    $0.902 \pm 0.004$ &  $0.194 \pm 0.005$ &                $1.227 \pm 0.003$ &  $0.177 \pm 0.004$ \\
$C_{A;\alpha+p1}$                                  &                    $0.755 \pm 0.005$ &  $0.166 \pm 0.007$ &                $1.052 \pm 0.005$ &  $0.179 \pm 0.007$ \\
$C_{A;p1+p2}$                                      &                    $0.829 \pm 0.004$ &  $0.181 \pm 0.005$ &                $1.131 \pm 0.005$ &  $0.166 \pm 0.007$ \\
$C_{A;p1}$                                         &                    $0.657 \pm 0.005$ &  $0.150 \pm 0.007$ &                $0.938 \pm 0.005$ &  $0.183 \pm 0.007$ \\}

\newcommand{\data}{
$C^{( \mathrm{Ani})}_{A;\alpha+p1+p2} $            &                    $0.869$ &  $0.177$ &                $1.167$ &  $0.169$ \\
$C^{( \mathrm{Ani})}_{A;\alpha+p1+p2} \; (p_{\tilde{v}}) $ &                    $0.999$ &  $0.244$ &                $1.339$ &  $0.256$ \\
$C^{( \mathrm{Ani})}_{A;\alpha+p1} $               &                    $0.730$ &  $0.142$ &                $0.997$ &  $0.156$ \\
$C^{( \mathrm{Ani})}_{A;\alpha+p1} \; (p_{\tilde{v}}) $  &                    $0.761$ &  $0.164$ &                $1.048$ &  $0.172$ \\
$C^{( \mathrm{Ani})}_{A;p1+p2} $*                   &                    $0.784$ &  $0.160$ &                \markprefered{$1.057$} &  \markprefered{$0.150$} \\
$C^{( \mathrm{Ani})}_{A;p1+p2} \; (p_{\tilde{v}}) $*  &                    $0.876$ &  $0.206$ &                $1.182$ &  $0.205$ \\
$C^{( \mathrm{Ani})}_{A;p1} $                      &                    \markprefered{$0.622$} &  \markprefered{$0.132$} &                $0.874$ &  $0.164$ \\
$C_{A;\alpha+p1+p2}$                               &                    $0.902$ &  $0.194$ &                $1.227$ &  $0.177$ \\
$C_{A;\alpha+p1}$                                  &                    $0.755$ &  $0.166$ &                $1.052$ &  $0.179$ \\
$C_{A;p1+p2}$*                                      &                    $0.829$ &  $0.181$ &                $1.131$ &  $0.166$ \\
$C_{A;p1}$                                         &                    $0.657$ &  $0.150$ &                $0.938$ &  $0.183$ \\}

\begin{deluxetable}{l|cc|cc}
\tablecaption{All fit parameters and their uncertainties in the manner calculated in Fig.~\ref{fig:ex-dvca-yAc-fits}. The column indicates the parameter (Mean Value or Width) for a given differentially flowing species. The row indicates the wave speed normalization. The bold, colored row is the preferred normalization. Anisotropic Alfv\'en speeds including the dynamic pressure term from Eq.~\ref{eq:CaAni} are indicated by $(p_{\tilde{v}})$. The average fit uncertainty on the $\mathrm{Mean}$ is $4\times 10^{-3}$ and the average uncertainty on $\mathrm{Width}$ is $5\times10^{-3}$. Normalizations marked with an asterisk (*) are plotted in Fig.~\ref{fig:mu-sigma}.\label{tbl:mu-sigma}}
\tablehead{
Wave Speed & \multicolumn{2}{c}{$\alpha-\mathrm{Particle}$}    & \multicolumn{2}{c}{Proton Beam}    \\
Normalization &    Mean &    Width &    Mean &    Width
}
    \startdata
%\datauncertainties
\data
\enddata
\end{deluxetable}

%% file: new_discussion.tex
\section{Discussion}
\label{sec:disc}

\added{The evolution of solar wind velocity distribution functions is governed by an interplay between adiabatic expansion, Coulomb collisions, and wave-particle interactions. Collisional transport rates \citep{Livi1986a, Pezzi2016} and many types of wave-particle interactions \citep{Verscharen2013,Verscharen2013a,Verscharen2013c} depend on the small-scale structure of the VDF, in particular the small-scale velocity space gradients. Because measurements indicate the presence of alpha-proton differential flow starting at the corona and extending out to and beyond 1 AU, one can assume that non-zero differential flow is a coronal signature. Under this hypothesis, the decay of $\dv{{\alpha,p1}}$ is due to dynamical friction. \citep{Kasper2017} As the proton beam-core drift and alpha-core drift are signatures of one plasma with a single expansion history, the collisional bottleneck that erodes $\dv{{\alpha,p1}}$ could likewise be expected to erode $\dv{{p2,p1}}$. However, the observed independence of $\dv{{p2,p1}}/C_A$ with respect to $A_c$ over the examined range contradicts this assumption and minimally implies either (1) an additional competing process that preferentially couples to proton beams or (2) that Eq. (\ref{eq:Ac}) underestimates the proton dynamical friction.}

\added{Several in situ mechanisms that preferentially couple to protons have been proposed. As one example, the interaction between resonant protons and kinetic Alfv\'en waves leads to the local formation of beams \citep{Voitenko2015}. Such a mechanism could be responsible for the creation of proton beams throughout the solar wind’s evolution or it could turn on at some distance from the sun where plasma conditions become favorable. As another example, \citet{Livi1987} have argued that Coulomb scattering itself in the presence of the interplanetary magnetic field can produce skewed and beam-like distributions under certain circumstances.}

\added{The collisional age used in Eq. (\ref{eq:Ac}) assumes that the collision frequency describing proton dynamical friction does not change over the solar wind’s evolution and is equal to the value measured at the spacecraft. \citet{Chhiber2016} have shown that such assumptions do not capture the full nature of proton radial evolution. Eq. (\ref{eq:Ac}) also neglects the ways in which this frequency depends on the small-scale structure of the VDF \citep{Livi1986a, Pezzi2016}. One avenue of future work is to better address collisional effects by modeling the radial dependence, building on the work of \citet{Chhiber2016} and \citet{Kasper2017}. A further refinement would be to account for dependence of collision frequency on the VDF fine structure \citep{Livi1986a, Pezzi2016}. A second avenue of future work involves modeling the force required to locally maintain differential flow. By letting this force depend on local wave amplitudes, perhaps the differential flow radial evolution could be modeled from the competition between a Coulomb frictional force and a force from resonant scattering \citep{Voitenko2015}.}

\added{The hypotheses of proton beams as coronal in origin or created and modified in situ are not mutually exclusive. For example, wave-resonant or frictional forcing may only be significant over a certain portion of the solar wind’s radial evolution and that range may correspond to a subset of commonly measured conditions at 1 AU. Applying a holistic model to data that is differentiated by wave power or Coulomb collisions may allow us to distinguish between or unite the two origin hypotheses. The upcoming Parker Solar Probe \citep{Fox2015} and Solar Orbiter \citep{Muller2013} missions, with their closer perihelia and higher energy resolution plasma instruments \citep{Kasper2015}, will also allow us to gauge the relative importance of and interplay between these effects.}